\documentclass[aps,showpacs,onecolumn,superscriptaddress]{revtex4-1}
\usepackage{amsmath}
\usepackage{amssymb}
\usepackage{graphicx}
\usepackage{xcolor}
\usepackage{times, txfonts}
\usepackage{bm}
\usepackage[bookmarks=false]{hyperref}
\hypersetup{colorlinks=true, citecolor=blue, linkcolor=blue}

\newcommand{\Tr}[1]{\mathrm{Tr}(\, #1)}
\newcommand{\C}[1]{\mathcal{C}_{#1}}

\begin{document}

\title{Single-shot realization of nonadiabatic holonomic gates with a superconducting Xmon qutrit}

\author{Zhenxing Zhang}
\affiliation{Physics Department, Zhejiang University, Hangzhou, 310027, China}
\author{P. Z. Zhao}
\affiliation{Department of Physics, Shandong University, Jinan 250100, China}
\author{Tenghui Wang}
 \affiliation{Physics Department, Zhejiang University, Hangzhou, 310027, China}
\author{Liang Xiang}
\affiliation{Physics Department, Zhejiang University, Hangzhou, 310027, China}
\author{Zhilong Jia}
\affiliation{Key Laboratory of Quantum Information, University of Science and Technology of China, Hefei, 230026, China}
\author{Peng Duan}
\affiliation{Key Laboratory of Quantum Information, University of Science and Technology of China, Hefei, 230026, China}
\author{D. M. Tong}
\email{tdm@sdu.edu.cn}
\affiliation{Department of Physics, Shandong University, Jinan 250100, China}
\author{Yi Yin}
\email{yiyin@zju.edu.cn}
\affiliation{Physics Department, Zhejiang University, Hangzhou, 310027, China}
\author{Guoping Guo}
\email{gpguo@ustc.edu.cn}
\affiliation{Key Laboratory of Quantum Information, University of Science and Technology of China, Hefei, 230026, China}

\date{\today}

\begin{abstract}
Nonadiabatic holonomic quantum computation has received increasing attention due to its robustness against
control errors as well as high-speed realization. The original protocol of nonadiabatic holonomic one-qubit gates
has been experimentally demonstrated with superconducting transmon qutrit.
However, the original protocol requires two noncommuting gates to realize an arbitrary one-qubit gate, which doubles
the exposure time of gates to error sources and therefore makes the gates vulnerable to environment-induced decoherence.
Single-shot protocol was subsequently proposed to realize an arbitrary one-qubit nonadiabatic holonomic gate.
In this paper, we experimentally realize the single-shot protocol of nonadiabatic holonomic single qubit gates with
a superconducting Xmon qutrit, where all the Clifford element gates are realized by a single-shot implementation.
Characterized by quantum process tomography and randomized benchmarking, the single-shot gates reach a fidelity larger
than 99\%.
\end{abstract}

\maketitle

\section{Introduction}

The circuit-based quantum computation requires a universal set of quantum gates,
including arbitrary one-qubit gates and a nontrivial two-qubit gate.
Geometric quantum computation is an interesting approach to implement the universal
quantum gates by using the
geometric phases \cite{Berry1984, WilczekPRL84, AharonovPRL87, AnandanPLA88}
or their non-Abelian counterpart, the holonomies \cite{ZanardiPLA99}.
Geometric phases are dependent on evolution paths but independent of evolution details, leading
to a build-in resilience to certain noises and control errors.
The early schemes of geometric quantum computation \cite{JonesNat00,ZanardiPLA99,DuanLMSci01} are based
on adiabatic geometric phases \cite{Berry1984,WilczekPRL84}. However, with the long operation time
required in an adiabatic process, the quantum gates are vulnerable to the environment-induced
decoherence in the system. To overcome this difficulty, nonadiabatic geometric quantum computation \cite{WangXBPRL01,ZhuSLPRL02}
based on nonadiabatic Abelian geometric phases \cite{AharonovPRL87} and nonadiabatic holonomic
quantum computation \cite{SjoqvistNJP12,XuGFPRL12} based on nonadiabatic non-Abelian geometric
phases \cite{AnandanPLA88} were proposed. The nonadiabatic holonomic quantum computation
retains the merits of both robustness against control errors and high-speed realization,
therefore receiving increasing attention for practical applications \cite{ZhangJPRA14,LiangZTPRA14,XueZYPRA15,ZhangJSciRep15,XuGFPRA15,SjoqvistPLA16,HerterichPRA16,WangYMPRA16,AbdumalikovNat13,
FengGRPRL13,ZuCNat14,CamejoNatCommu14,ZhouBBPRL17,SekiguchiNatPhoton17,LiHSCPMA17,XuYPRL18,HongZPPRA18}.

The original protocol \cite{SjoqvistNJP12,XuGFPRL12} of nonadiabatic holonomic one-qubit
gates has been experimentally demonstrated with a superconducting circuit \cite{AbdumalikovNat13},
nuclear magnetic resonance \cite{FengGRPRL13}, and nitrogen-vacancy centers in diamond \cite{ZuCNat14,CamejoNatCommu14}.
However, based on the original protocol, a single step operation can only rotate the quantum state about arbitrary axes, with a fixed angle of $\pi$.
An arbitrary one-qubit gate then requires two sequential
steps, which doubles the exposure time of gates to error sources.
A single-shot protocol of nonadiabatic holonomic gates is further proposed, in which
the quantum state is rotated about arbitrary axes with variable angles \cite{XuGFPRA15,SjoqvistPLA16}.
Recently, the single-shot protocol of nonadiabatic holonomic gates were experimentally realized
with nitrogen-vacancy centers in diamond \cite{ZhouBBPRL17,SekiguchiNatPhoton17} and nuclear magnetic resonance \cite{LiHSCPMA17}.
With a different approach, a single-loop protocol of holonomic gates is also proposed \cite{HerterichPRA16, HongZPPRA18} and experimentally realized \cite{XuYPRL18}.

A superconducting circuit provides an appealing scalable platform for implementing nonadiabatic
holonomic quantum computation. As a solid state system, the integrated circuit can be easily
scaled up to a multi-qubit system, with each qubit controlled by individual lines.
The superconducting Xmon is a high quality qubit with relative long coherence time, the design of which
also balances the coherence, the connectivity, as well as the fast control \cite{BarendsPRL13,BarendsNat14,KellyNat15}.
In this paper, we experimentally realize the single-shot protocol of nonadiabatic holonomic
one-qubit gates with a three-level superconducting Xmon qutrit.

An arbitrary single qubit gate can be realized with the single-shot protocol.
Here we choose a group of single qubit Clifford gates as examples to present results.
Excluding the identity operation, we classify the
group to $\pi$-rotation, $\pi/2$-rotation, and $2\pi/3$-rotation Clifford gates.
The $\pi$-rotation gates are simultaneously driven by two resonant microwave pulses as in the
original protocol \cite{SjoqvistNJP12,XuGFPRL12}, while the other gates are driven by two off-resonant pulses following the single-shot protocol \cite{XuGFPRA15,SjoqvistPLA16}.
By using both the quantum process tomography (QPT) \cite{NielsenChuang} and randomized benchmarking (RB) \cite{ChowPRL09,MagesanPRL11,MagesanPRL12},
we demonstrate the realization of the single-shot gate with high fidelity.

\section{Results}

\textbf{Protocol}

We first explain how a nonadiabatic holonomic gate arises \cite{SjoqvistNJP12,XuGFPRL12}.
Consider a quantum system described by a $N$-dimensional state space and exposed to the Hamiltonian $H(t)$.
Assume there is a time-dependent $L$-dimensional subspace
$\mathcal{S}(t)=\mathrm{Span}\{|\psi_{k}(t)\rangle\}^{L}_{k=1}$, where $|\psi_{k}(t)\rangle$
satisfies the Schr\"{o}dinger equation $i|\dot{\psi}_{k}(t)\rangle=H(t)|\psi_{k}(t)\rangle$.
We take $\mathcal{S}(0)$ as the computational space.
If the following requirements
\begin{align}
&(\mathrm{i})~\sum^{L}_{k=1}|\psi_{k}(\tau)\rangle\langle\psi_{k}(\tau)|
=\sum^{L}_{k=1}|\psi_{k}(0)\rangle\langle\psi_{k}(0)|, \notag\\
&(\mathrm{ii})~~\langle\psi_{k}(t)|H(t)|\psi_{l}(t)\rangle=0,~k,l=1,2,\cdot\cdot\cdot,L,
\end{align}
are satisfied, the unitary transformation acting on $\mathcal{S}(0)$ is a nonadiabatic holonomic gate.
Here, condition $(\mathrm{i})$ entails that $\mathcal{S}(t)$ undergoes a cyclic evolution
and condition $(\mathrm{ii})$ ensures a parallel transport with vanishing dynamical phases.

\begin{figure}[htbp]
   \includegraphics[width=0.5\linewidth]{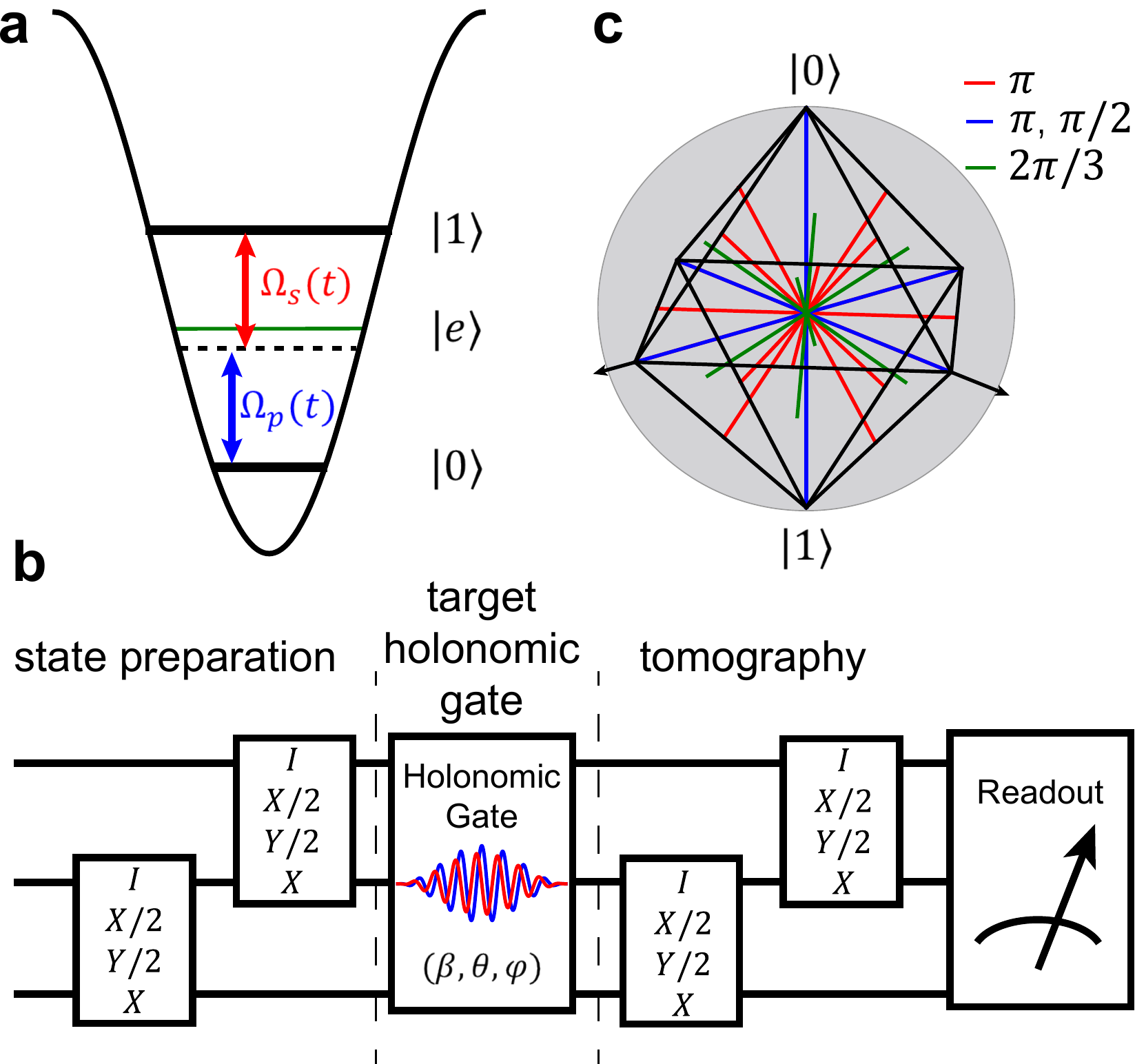}
   \caption{
   Single-shot protocol single-qubit holonomic gates.
   (a) Setup for cascaded configuration of the three-level Xmon qutrit. $|0\rangle$, $|e\rangle$ and $|1\rangle$ are three lowest levels
   of the superconducting circuit. (b) Schematic diagram of QPT measurement for a holonomic gate. (c) Octahedral group and its rotation axes. The red, blue, and green color lines represent the axis of $\pi$-, $\pi/2$- and $2\pi/3$-rotation respectively. The $x$-, $y$-, and $z$-axis in blue are also the axis of $\pi$-rotation.}
   \label{Fig1}
\end{figure}

In our experiment, we realize a single-shot protocol of nonadiabatic holonomic gate \cite{XuGFPRA15,SjoqvistPLA16}.
Consider a three-level superconducting Xmon consisting of three lowest levels $|0\rangle$,
$|e\rangle$ and $|1\rangle$ with ladder configuration, as shown in Fig.~\ref{Fig1}(a).
The states $|0\rangle$ and $|1\rangle$ are taken as the qubit computational basis while the state $|e\rangle$
acts as an auxiliary state, with $\omega_{oe}$ and $\omega_{e1}$ being energy differences
between neighboring states.
The transitions $|0\rangle\leftrightarrow|e\rangle$ and $|e\rangle\leftrightarrow|1\rangle$
are facilitated by two microwave pulses with pump frequency $\omega_{p}(t)$
and stocks frequency $\omega_{s}(t)$ \cite{VitanovRMP17}. Here the two microwave drives are off-resonant with detunings
$\Delta_{p}(t)$ and $\Delta_{s}(t)$, where $\Delta_{p}(t)=\omega_{oe}-\omega_{p}(t)$ and $\Delta_{s}(t)=\omega_{1e}-\omega_{s}(t)$.
In the two-photon resonance condition $\omega_{p}(t)+\omega_{s}(t)=\omega_{oe}+\omega_{e1}$,
the Hamiltonian of the Xmon, in the double-rotating frame and by using rotating wave
approximation, reads
\begin{align}
H(t)=\Delta_{p}(t)|e\rangle\langle e|
+\frac{1}{2}\left[\Omega_{p}(t)|e\rangle\langle0|+\Omega_{s}(t)|e\rangle\langle1|+\mathrm{H.c.}\right],
\label{eq1}
\end{align}
where $\Omega_{p}(t)$ and $\Omega_{s}(t)$ are time-dependent envelopes and
$\mathrm{H.c.}$ represents the Hermitian conjugate terms.

To realize the nonadiabatic holonomic gates, the parameters in Eq.~(\ref{eq1}) are taken as
\begin{align}
\Delta_{p}(t)&=\Omega(t)\sin\alpha,\notag\\
\Omega_{p}(t)&=\Omega(t)\cos\alpha\cos\frac{\theta}{2},\notag\\
\Omega_{s}(t)&=\Omega(t)\cos\alpha\sin\frac{\theta}{2}e^{-i\varphi},
\end{align}
where $\Omega(t)$ is time-dependent, and $\alpha$, $\theta$ and $\varphi$ are time-independent constants.
Accordingly, the Hamiltonian in Eq.~(\ref{eq1}) can be expressed as
\begin{align}
\mathcal{H}(t)=\Omega(t)\sin\alpha|e\rangle\langle e|+\frac{1}{2}\Omega(t)\cos\alpha(|e\rangle\langle b|+|b\rangle\langle e|),
\end{align}
with the bright state $|b\rangle=\cos\frac{\theta}{2}|0\rangle+\sin\frac{\theta}{2}e^{i\varphi}|1\rangle$.
Here, a dark state is orthogonal to the bright state, as $|d\rangle=\sin\frac{\theta}{2}e^{i\varphi}|0\rangle-\cos\frac{\theta}{2}|1\rangle$.
If the evolution period $\tau$ is taken to satisfy
\begin{align}
\int^{\tau}_{0}\frac{\Omega(t)}{2}dt=\pi, \label{eq2}
\end{align}
the evolution operator can be obtained as
\begin{align}
U(\tau)=|d\rangle\langle d|+e^{-i\pi(1+\sin\alpha)}|b\rangle\langle b|+e^{-i\pi(1+\sin\alpha)}|e\rangle\langle e|.
\end{align}
Consequently, an arbitrary one-qubit gate can be obtained by projecting the evolution operator onto the computational subspace, and it is
\begin{align}
U_{L}(\tau)=e^{-i\gamma\bm{n}\cdot\bm{\sigma}/2},
\label{eq3}
\end{align}
where $\gamma=\pi(1+\sin\alpha)$, $\bm{n}=(\sin\theta\cos\varphi,\sin\theta\sin\varphi,\cos\theta)$,
and $\bm{\sigma}=(\sigma_{x},\sigma_{y},\sigma_{z})$ with $\sigma_{x},\sigma_{y},\sigma_{z}$
being the Pauli operators acting on computational basis $|0\rangle$ and $|1\rangle$.
Here, an unimportant global phase is neglected.

We now demonstrate that $U_{L}(\tau)$ is a nonadiabatic holonomic gate.
First, condition (i) is satisfied since $U(\tau)(|0\rangle\langle0|+|1\rangle\langle1|)U^{\dagger}(\tau)=|0\rangle\langle0|+|1\rangle\langle1|$.
Second, with the aid of  the commutation relation $[\mathcal{H}(t),U(t)]=0$, we
can verify that condition (ii) is satisfied since $\langle\Phi_{i}(t)|\mathcal{H}(t)|\Phi_{j}(t)\rangle=\langle i|U^{\dagger}(t)\mathcal{H}(t)U(t)|j\rangle
=\langle i|\mathcal{H}(t)|j\rangle=0$, where $|\Phi_{i(j)}(t)\rangle=U(t)|i(j)\rangle$
with $U(t)=e^{-i\int^{t}_{0}\mathcal{H}(t^{\prime})dt^{\prime}}$ and $i(j)=0,1$.
Therefore, $U_{L}(\tau)$ is a nonadiabatic holonomic gate.

In the original protocol of nonadiabatic holonomic gates, the three level system is controlled by two resonant pulses with $\Delta_p(t)=0$, which is
equivalent to a fixed rotation angle of $\gamma=\pi$. In the single-shot protocol, the two off-resonant pulses are applied with a variable detuning.
An arbitrary rotation angle $\gamma$ can be obtained. In other words, an arbitrary one-qubit gate can be realized in a single-shot implementation.
The original protocol is a specific case of the current single-shot protocol.

\textbf{Single Qubit Clifford Group}

An arbitrary single qubit gate can be implemented with the single-shot nonadiabatic holonomic protocol,
by choosing specific gate parameters $\alpha$, $\theta$ and $\varphi$.
To demonstrate the arbitrariness of the protocol, we implement single qubit gates in the Clifford group (Clifford gates).
For a single qubit, Clifford gates consist 24 rotations preserving quantum states along vertexes of an octahedron in the Bloch sphere.
The rotation axes are lines connecting the origin of the Bloch sphere and a face center, vertex or midpoint of an edge of the octahedron, as shown in Fig.~\ref{Fig1}(c).
Classified by the rotation angle, the single qubit Clifford gates can be divided into four sets: identity, $\pi$-rotation, $\pi/2$-rotation and $2\pi/3$-rotation.
For the set of $\pi$-rotation, there are nine single qubit gates in the Clifford group, red axis and blue axis in Fig.~\ref{Fig1}(c) . This set of gates can be realized by the original
nonadiabatic holonomic protocol, and implemented with a single step of resonant pump and stocks drives.
However, for the sets of $\pi/2$- (blue axis) and $2\pi/3$- (green axis) rotations, the remaining 14 gates have to be realized by combing two $\pi$-rotation gates in the original
nonadiabatic holonomic protocol. With the single-shot protocol, these two sets can be implemented with a single step of off-resonant
pump and stocks drives.

\textbf{Experimental Parameters}

The superconducting Xmon used in this work is an aluminum-based circuit operated at about $10$ mK in
a cryogen-free dilution refrigerator. For a single Xmon in this experiment, the lowest three levels
are used as $|0\rangle$, $|e\rangle$, and $|1\rangle$ in the single-shot protocol. The relevant transition
frequencies are $\omega_{0e}/{2\pi}=4.849$ GHz and $\omega_{e1}/{2\pi}=4.597$ GHz, and the nonlinearity
$\eta = (\omega_{e1}-\omega_{0e})/{2\pi}=-252$ MHz. The coherence of the qubit is characterized by
energy relaxation time $T_1^{e0} = 25.3\mu$s, $T_1^{1e}=12.8\mu$s, and the pure dephasing time
$T_\phi^{e0}=28.1\mu$s, $T_\phi^{1e}=13.4\mu$s measured with a Ramsey interference experiment.

In the single-shot nonadiabatic holonomic protocol, the rotation angle $\gamma$ is determined by the time-independent
parameter $\alpha$. For the single qubit Clifford gates, the set of $\gamma=\pi$, $\pi/2$- and $2\pi/3$- rotations
are accordingly assigned with $\alpha=0$, $-\pi/6$ and $-\arcsin(1/3)$. The rotation axis for each gate is
determined by two other time-independent parameters $\theta$ and $\varphi$.
The same time-dependent pulse envelope $\Omega(t)$ is shared for all the gates regardless of different gate parameters $\beta$, $\theta$ and $\varphi$.
The form of $\Omega(t)$ is designed as $\Omega(t)=\Omega_0\sin^2(\pi t/\tau)$, with the gate time $\tau$ set at 100 ns and the constraint
of Eq.~(\ref{eq2}) as $\Omega_0 \tau = 4\pi$. For this specified form of $\Omega(t)$, the microwave drive pulse is turned on and off smoothly with $\Omega(0)=\Omega(\tau)=0$.
Following we present the realization of all single qubit Clifford gates with the nonadiabatic holonomic single-shot protocol.

\begin{figure}[htbp]
\centering
\includegraphics[width=0.5\linewidth]{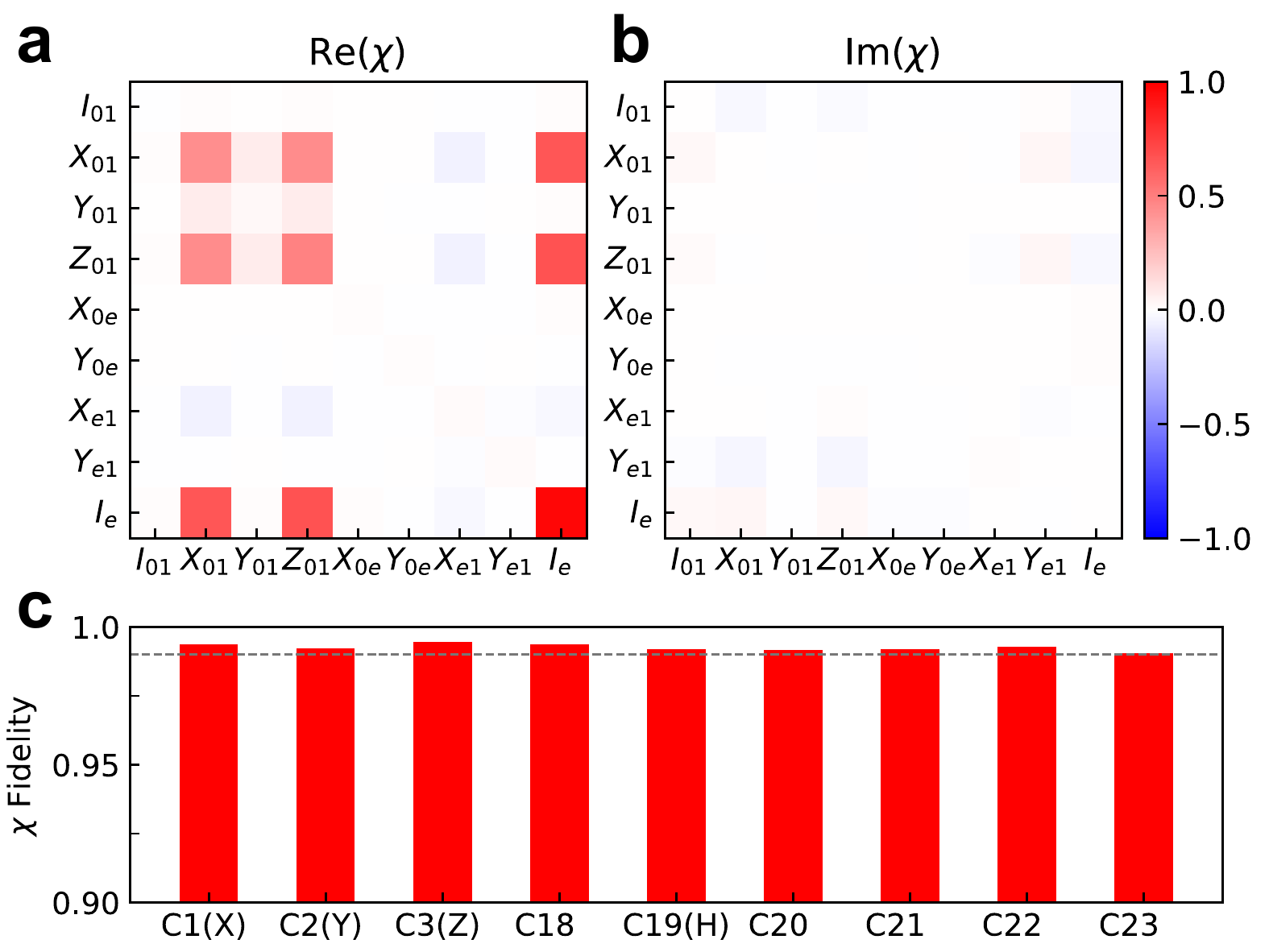}
\caption{QPT result and the fidelities. The real (a) and the imaginary (b) parts of process matrix $\chi$ for $\pi$-rotation. (c) Fidelities for all $\pi$-rotation gates.
The dashed line represents a fidelity of 99\%.}
\label{Fig2}
\end{figure}

\textbf{$\pi$-rotation Clifford gates}

We initially implement holonomic gates with $\pi$-rotation in the Clifford group.
In this special case, the single-shot protocol falls back to the original protocol.
Accordingly, two resonant microwave pulses simultaneously drive the
$|0\rangle\leftrightarrow|e\rangle$ and $|e\rangle\leftrightarrow|1\rangle$ level transitions with the pump pulse
$\Omega_p(t)=\Omega(t)\cos\frac{\theta}{2}$ and the drive pulse $\Omega_s(t)=\Omega(t)\sin\frac{\theta}{2}$, respectively.
The gate parameter $\alpha=0$ (i.e. $\Delta_p(t)=0$),
yielding the rotation angle $\gamma=\pi$.
The rotation axis of the quantum gates is determined by choosing specific parameters $\theta$ and $\varphi$.
To completely characterize the holonomic gates, we perform a quantum process tomography (QPT) involving
all three basis $|0\rangle, |e\rangle, |1\rangle$ \cite{BianchettiPRL10}, which is quantified
by a reconstructed $\chi$-matrix.
As shown in Fig.~\ref{Fig1}(b), for the 3-level QPT, we prepare 16 different initial states $\rho_i$
by sequentially applying the identity $I$, the $\pi/2$-rotation $X/2$, $Y/2$ and the $\pi$-rotation $X$ pulses to $|0\rangle\leftrightarrow|e\rangle$ and $|e\rangle\leftrightarrow|1\rangle$ transitions.
Initialized states are then followed by a specified holonomic quantum gate in Clifford group.
Finally we perform a state measurement with full quantum state tomography (QST).
The output state $\rho_f$ is extracted using the maximum likelihood estimation method.
The process matrix, $\chi$, is reconstructed from the input and output states by numerically solving the equation
$\rho_f = \sum_{m, n} \chi_{mn} E_m \rho_i E_n^\dagger$ \cite{NielsenChuang}.
The full set of nine orthogonal basis operators $E_m$ is chosen as $\{I_{01}, X_{01}, Y_{01}, Z_{01}, X_{0e}, Y_{0e}, X_{e1}, Y_{e1}, I_e\}$ \cite{BianchettiPRL10,AbdumalikovNat13}.
The first four operators represent the operation between the computation subspace, $\{|0\rangle, |1\rangle\}$.
For the holonomic gate within the subspace, a population leakage to the auxiliary state $|e\rangle$ may happen due to
nonideal microwave pulses, and energy relaxation/dephasing of both $|e\rangle$ and $|1\rangle$.
The leakage can be determined by the trace of the reduced $\chi$-matrix, $\tilde{\chi}$, which describes the process involving the computation states $|0\rangle$ and $|1\rangle$ \cite{AbdumalikovNat13}.

As an example of the $\pi$-rotation gate, a Hadamard gate ($H$) is experimentally realized and demonstrated, with the
setting parameters $\alpha=0$, $\theta=\pi/4$ and $\varphi=0$.
The real and imaginary parts of $\chi$-matrix for the Hadamard gate are shown in Fig.~\ref{Fig2}(a) and Fig.~\ref{Fig2}(b), respectively.
The dominant elements in the subspace $\{|0\rangle, |1\rangle\}$ is $X_{01}$ and $Z_{01}$, and
the imaginary part of $\chi$-matrix is close to zero, which are both the same as the theoretical expectation.
Moreover, the element $\chi_{I_e, I_e}$ is close to one, which represents that the auxiliary state is nearly unaffected during the holonomic gate operation, as described in the single-shot protocol.
The population leakage of Hadamard gate is described by the trace of the reduced $\chi$-matrix,
$\Tr{\tilde{\chi}}$, which is about 0.96.
The main leakage error comes from the imperfection of the microwave signal and the limited nonlinearity.
Following the definition of fidelity \cite{FengGRPRL13,WangXGPLA08,ZhangJFPRL12},
\begin{equation}
F = \frac{|\Tr{(\chi_{\mathrm{exp}}\chi_{\mathrm{th}}^\dagger)}|}
{\sqrt{\Tr{(\chi_\mathrm{exp}\chi_{\mathrm{exp}}^\dagger)}\Tr{(\chi_\mathrm{th}\chi_\mathrm{th}^\dagger)}}},
\end{equation}
we calculate the fidelity of the process matrix using the ideal process matrix $\chi_\mathrm{th}$ as a reference.
For the Hadamard gate, the fidelity of QPT $F(H)$ reaches 99.2\%,
and the error of the process is less than 1\%.
In Fig.~\ref{Fig2}(c), we show the fidelity for all $\pi$-rotation gates in single qubit Clifford group with specific gate parameters $\theta$ and $\varphi$.
The fidelities for all these gates are above 99\%, with an average fidelity 99.3\%.

\begin{figure}[htbp]
\centering
\includegraphics[width=0.5\linewidth]{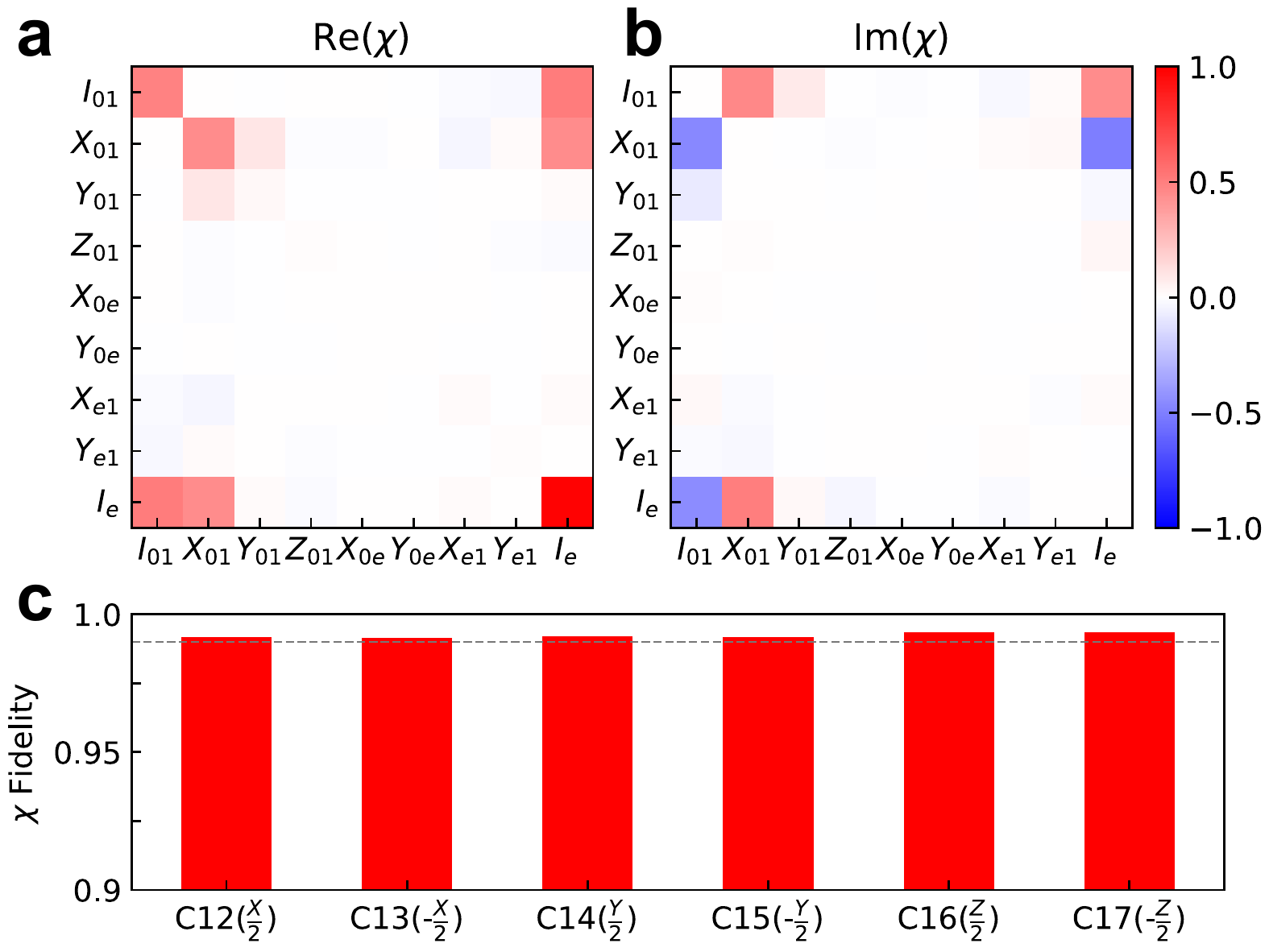}
\caption{QPT result and the fidelities. The real (a) and the imaginary (b) parts of process matrix $\chi$ for $\pi/2$-rotation. (c) Fidelities for all $\pi/2$-rotation gates.
The dashed line represents a fidelity of 99\%.}
\label{Fig3}
\end{figure}

\textbf{$\pi/2$-rotation Clifford gates}

To go beyond the $\pi$-rotation gates, we follow the single-shot protocol and implement
holonomic gates with $\pi/2$-rotation in the Clifford group.
To obtain a rotation angle $\gamma=\pi/2$, the off-resonant pump pulse and stocks pulse are simultaneously applied,
with the parameter $\alpha$ set as $-\pi/6$ and the detuning $\Delta_p(t) = -\Omega(t)/2$.
Different $\pi/2$-rotation Clifford gates are specified by their corresponding rotation axes,
which determine the parameters $\theta$ and $\varphi$.

As an example of $\pi/2$-rotation gates, following we realize and demonstrate
the $X/2$ gate by setting $\alpha=-\pi/6$, $\theta=\pi/2$ and $\varphi=0$.
With the same QPT measurement as described previously, we reconstruct the process $\chi$-matrix
for $X/2$ gate and present its real and imaginary parts in Fig.~\ref{Fig3}(a) and Fig.~\ref{Fig3}(b), respectively.
The $\chi$-matrix of $X/2$ includes auto and cross correlations between operators $I$ and $X$.
The element of $I_e$, $\chi_{I_e, I_e}$, is close to 1, showing neglectable effect to the auxiliary state  during the gate operation.
The population leakage to the auxiliary subspace is about 0.97, characterized by $\Tr{\tilde{\chi}}$.
The fidelity of the full QPT is $F(\frac{X}{2})=99.2\%$, which is similar to the fidelity in $\pi$-rotation gates.
In Fig.~\ref{Fig3}(c), we show the fidelities of all the gates with $\pi/2$-rotation in single qubit Clifford group.
From the experimental result, all the fidelities are above 99\% with an average fidelity 99.2\%.

With the original protocol of holonomic gates, a $\pi/2$-rotation gate can only be realized by sequentially combining
two $\pi$-rotation gates. The fidelity of such a gate can be roughly estimated to be about 98\%, with a similar
fidelity of $\pi$-rotation gate as in our experiment. With the single-shot protocol, the arbitrary gate can be
implemented within a single step. By shortening the gate operation time, we reduce the accumulated
environment-induced error in a combined operation, and increase the corresponding fidelity.

\begin{figure}[htbp]
\centering
\includegraphics[width=0.5\linewidth]{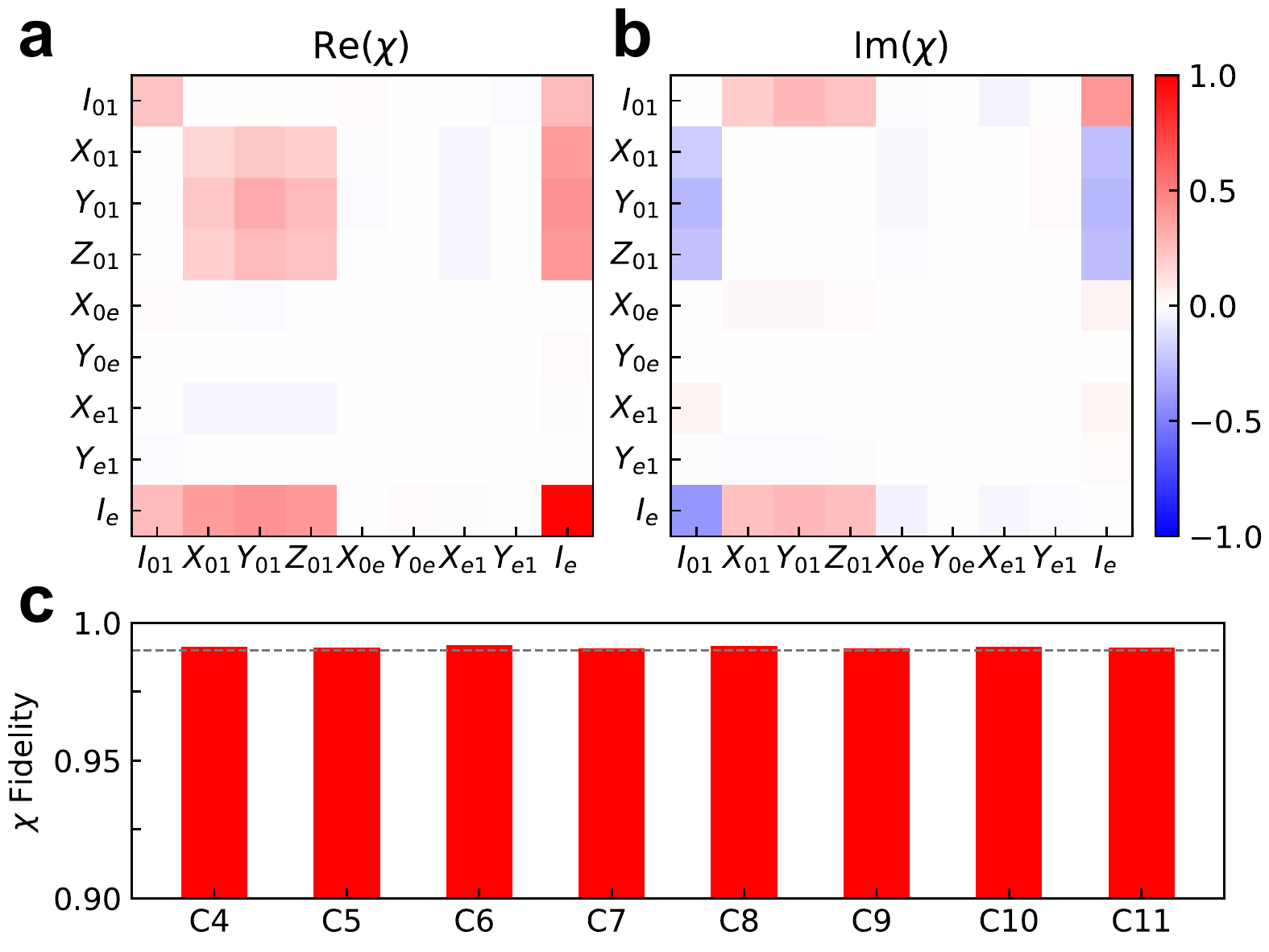}
\caption{QPT result and the fidelities. The real (a) and the imaginary (b) parts of process matrix $\chi$ for $2\pi/3$-rotation. (c) Fidelities for all $2\pi/3$-rotation gates.
The dashed line represents a fidelity of 99\%.}
\label{Fig4}
\end{figure}

\textbf{$2\pi/3$-rotation Clifford gates}

To complete the whole Clifford group of single-qubit gates, we realize and demonstrate
eight $2\pi/3$-rotation gates with the single-shot protocol. The axes of these gates
are the lines connecting the origin and face centers of the octahedron, as shown in Fig.~\ref{Fig1}(c).
For all the $2\pi/3$-rotation gates, we set $\alpha=-\arcsin(1/3)$, leading to a rotation angle $\gamma=2\pi/3$.
The detuning satisfies $\Delta_p(t) = -\Omega(t)/3$, and the other two parameters $\theta$
and $\varphi$ are determined by the corresponding rotation axis for each gate.

For example, the gate $\mathcal{C}_8$ is an operation of $2\pi/3$-rotation about the axis $\bm{n}=\frac{1}{\sqrt{3}}(1,1,1)$,
which determine the parameters $\theta=\arccos(1/3)$ and $\varphi=\pi/4$. From Eq.~(\ref{eq3}), the gate
operation in the computation subspace for $\mathcal{C}_8$ reads as:
\begin{equation}\label{eq9}
U=\frac{1}{2}\begin{bmatrix}1-i& -1-i\\1-i& 1+i\\ \end{bmatrix}
=\frac{1}{2}(I-i\sigma_x -i \sigma_y-i\sigma_z).
\end{equation}
In Fig.~\ref{Fig4}(a) and Fig.~\ref{Fig4}(b), we show the $\chi$-matrix for the gate $\mathcal{C}_8$.
The real part of $\chi$-matrix shows similar value in the $X$, $Y$, $Z$ components,
agreeing with the theoretic expectation in Eq.~(\ref{eq9}).
The QPT fidelity for the gate $\mathcal{C}_8$ is about $F(\mathcal{C}_8)=99.2\%$, which is comparable to the previous holonomic gates
with other rotation angles.
And the population leakage characterized by the quantity $\Tr{\tilde{\chi}}$  is about 98\%.
In Fig.~\ref{Fig4}(c), we present the fidelities for all the Clifford gates with a $2\pi/3$-rotation.
All the fidelities are larger than 99\%, with an average fidelity reaching 99.1\%.

\begin{figure}[htbp]
\centering
\includegraphics[width=0.5\linewidth]{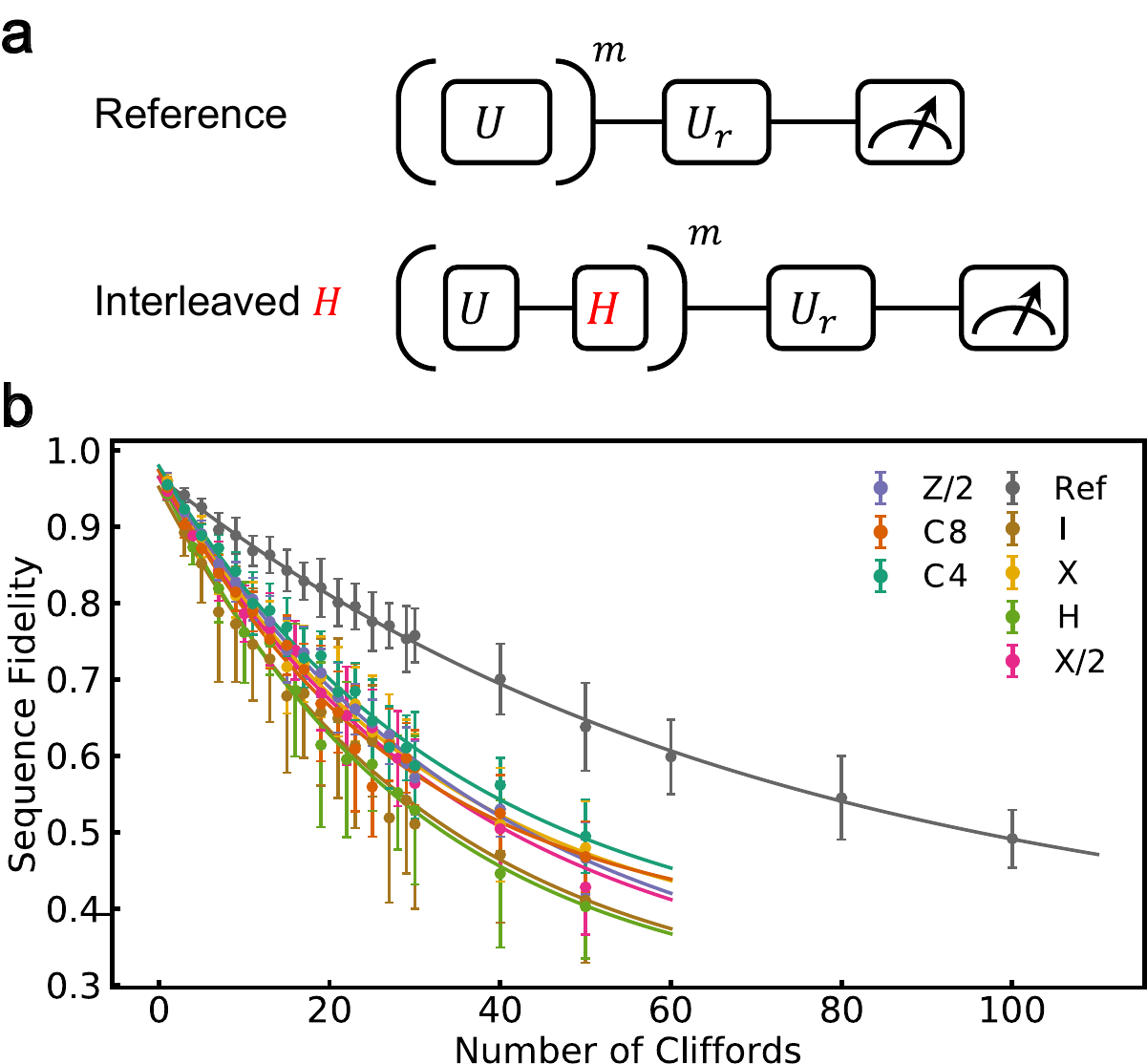}
\caption{RB of the single-qubit holonomic gates.
(a) The schematic diagram for reference and interleaved randomized benchmarking measurement.
(b) The sequence fidelities are displaced as functions of the number of Cliffords. Each sequence is average over $k=50$
randmized operation, with its standard deviation from the mean shown as an errorbar.}
\label{Fig5}
\end{figure}

\textbf{Randomized Benchmarking}

Randomized benchmarking (RB) is another systematic method to extract the quantum gate fidelity.
The gate fidelity in the RB measurement is separately quantified by
excluding errors in state preparation and measurement \cite{ChowPRL09,MagesanPRL11,MagesanPRL12,BarendsNat14}.
In this section, we perform the Clifford-based RB measurement to obtain the holonomic gate fidelity.
Twenty-four Clifford holonomic gates are used in the RB experiment. Each gate is
individually implemented by choosing specific gate parameters $\alpha$, $\theta$, and $\varphi$.

A reference RB experiment is performed first. As shown in the pulse sequence in Fig.~\ref{Fig5}(a),
the qubit is initially prepared at $|0\rangle$ state, then a sequence of $m$ Clifford gates
are randomly chosen to drive the qubit. Since the Clifford group is a closed set, a recovery
gate can be defined and finally applied to reverse the operation of $m$ Clifford gates.
The remaining population of the initial state is measured afterwards. After repeating this
random operation sequence $k$ ($=50$ in our experiment) times, we obtain the average result of
remaining population as a function of $m$, which is also called a sequence fidelity.
As shown in Fig.~\ref{Fig5}(b), the sequence fidelity can be fitted using the function \cite{MagesanPRL11}:
\begin{equation}F = A p^m + B\end{equation}
where $F$ is the sequence fidelity, $p$ is a depolarizing parameter, and the parameters $A$ and $B$
absorb the error in state preparation and measurement. The average error $r$ over the
randomized Clifford gates is given by $r=(1-p)/2$. From the fitting of the reference RB measurement,
we obtain the depolarized parameter $p_\mathrm{ref}=0.986$, which yields an average error $r_\mathrm{ref}=0.007$, or
an average RB fidelity 99.3\%.

The reference RB experiment can only give an average gate fidelity over the Clifford gates.
For a specific gate, interleaved RB experiment can be applied to determine the gate fidelity \cite{MagesanPRL12}.
In Fig.~\ref{Fig5}(a), we show the pulse sequence for the interleaved RB experiment.
At each step, the qubit is driven by a combination of a randomly selected Clifford gate and
the target holonomic gate. After a reversed recovery gate is applied, the remaining population
or the sequence fidelity is measured as a function of the number of steps $m$.
Similarly, the sequence fidelity is fitted by the same function as in the reference RB experiment,
leading to a new depolarized parameter $p_\mathrm{gate}$.

The specific gate fidelity is given by $F_\mathrm{gate} = 1 - (1-p_\mathrm{gate}/p_\mathrm{ref})/2$.
The experimental result for 7 gates are shown in Fig.~\ref{Fig5}(b). The chosen gates are the idle gate $I$, two $\pi$-rotation gates, $X$ and $H$,
two $\pi/2$-rotation gates $X/2$ and $Z/2$, and two $2\pi/3$-rotation gates $\mathcal{C}_8$ (rotation axis $\bm{n}=\tfrac{1}{\sqrt{3}}(1,1,1)$) and $\mathcal{C}_4$ (rotation axis $\bm{n}=\tfrac{1}{\sqrt{3}}(1,1,-1)$).
From the interleaved RB experiment, the fidelities for $X$, $H$, $X/2$, $Z/2$, $\mathcal{C}_8$ and $\mathcal{C}_4$ are
99.2\%, 99.0\%, 99.3\%, 99.4\%, 99.0\% and 99.3\%, respectively. After benchmarking all the Clifford gates, we
obtain RB fidelities for all the Clifford holonomic gates, each of which is larger than 99.0\%.

\section{Discussion}

From the QPT and RB experiments, we verify that all the single qubit Clifford gates can be implemented by single-shot protocol with high fidelity.
It is confirmed that all the fidelities for single qubit gates are higher than 99.0\%. We suggest that the main error comes from the decoherence and the energy relaxation, which is confirmed by the fidelity of the idle gate, $F=99.1\%$.
We obtain similar fidelities for $\pi$-rotation gates and gates with other rotation angles,
which means that the control errors are similar for resonant and off-resonant pulses.

The single qubit gates in our experiment are compatible with the previously proposed two-qubit nonadiabatic holonomic gate.
By combining two-qubit gates with the current results, we can obtain a universsal nonadiabatic holonomic quantum computation in the future.

\section{Methods}

The Xmon sample is fabricated on a silicon substrate, with a standard nano-fabrication method. Four arms of
the Xmon cross are connected to different lines for separate functions of coupling, control and readout.
The lowest three energy levels of the Xmon are utilized in the single-shot protocol. A readout resonator
couples the Xmon and a readout line for a dispersive measurement of the quantum state of the three-level qutrit.
A $1\,\mu$s-long microwave signal is send to the sample, with frequency $f=6.56$ GHz.
After interacting with the readout resonator, the signal is amplified by a Josephson parametric
amplifier \cite{RoyAPL15,YuanXPRL16} and a high electron mobility transistor (HEMT). The signal is further digitalized and demodulated
by an analog to digital convertor for a high fidelity measurement.
By heralding the ground state $|0\rangle$ \cite{JohnsonPRL12}, the readout fidelity for the lowest three levels
are $F_0=99.5\%$, $F_e=92.3\%$ and $F_1=89.5\%$, respectively.

The operations of gates $\mathcal{C}_i$ shown in the figures from Fig.~\ref{Fig2} to Fig.~\ref{Fig5}
are given in the supplementary information.

\section{Acknowledgement}

We thank L. Sun for providing us with the Josephson Parametric Amplifier (JPA) used in this work. The work was supported by the National Basic Research Program of China (2014CB921203, 2015CB921004),
the National Key Research and Development Program of China (2016YFA0301700, 2017YFA0304303),
the National Natural Science Foundation of China (NSFC-21573195, 11625419, 11474177),
the Fundamental Research Funds for the Central Universities in China, and the Anhui Initiative in Quantum Information Technologies (AHY080000).
This work was partially carried out at the University of Science and Technology of China Center for Micro and Nanoscale Research and Fabrication.

\clearpage

\section{Clifford Gates}
\begin{table}
   \centering
   \caption{The 23 single qubit Cliffords with the name $\C{i}$ and the rotation axis $\bm{n}$ excluding
   the identity operation $I$ ($\mathcal{C}_0$). Here $X/2$ donates a $\pi/2$ rotation over the X axis with unitary $R_X(\pi/2)=\exp(-i\pi\sigma_X/4)$.}
   \label{tab1}
   \begin{tabular}{p{3 cm}p{2.5 cm}p{2.5 cm}}
   \hline\hline
   & gate name & rotation axis ($\bm{n}$)\\
   \hline
   $\pi$-rotation
   & $\C{1}\; (X)$ & (1, 0, 0)\\
   & $\C{2}\; (Y)$ & (0, 1, 0)\\
   & $\C{3}\; (Z)$ & (0, 0, 1)\\
   & $\C{18}$      & $\tfrac{1}{\sqrt{2}}(1, 0, -1)$\\
   & $\C{19}\; (H)$  & $\tfrac{1}{\sqrt{2}}(1, 0, 1)$\\
   & $\C{20}$ & $\tfrac{1}{\sqrt{2}}(0, 1, 1)$\\
   & $\C{21}$ & $\tfrac{1}{\sqrt{2}}(0, 1, -1)$\\
   & $\C{22}$ & $\tfrac{1}{\sqrt{2}}(1, 1, 0)$\\
   & $\C{23}$ & $\tfrac{1}{\sqrt{2}}(1, -1, 0)$\\
   \hline
   $2\pi/3$-rotation
   & $\C{4}$ & $\tfrac{1}{\sqrt{3}}(1, 1, -1)$\\
   & $\C{5}$ & $\tfrac{1}{\sqrt{3}}(1, -1, 1)$\\
   & $\C{6}$ & $\tfrac{1}{\sqrt{3}}(-1, 1, 1)$\\
   & $\C{7}$ & $\tfrac{1}{\sqrt{3}}(-1, -1, -1)$\\
   & $\C{8}$ & $\tfrac{1}{\sqrt{3}}(1, 1, 1)$\\
   & $\C{9}$ & $\tfrac{1}{\sqrt{3}}(-1, 1, -1)$\\
   & $\C{10}$ & $\tfrac{1}{\sqrt{3}}(1, -1, -1)$\\
   & $\C{11}$ & $\tfrac{1}{\sqrt{3}}(-1, -1, 1)$\\
   \hline
   $\pi/2$-rotation
   & $\C{12}\; (X/2)$ & (1, 0, 0)\\
   & $\C{13}\; (-X/2)$ & (-1, 0, 0)\\
   & $\C{14}\; (Y/2)$ & (0, 1, 0)\\
   & $\C{15}\; (-Y/2)$ & (0, -1, 0)\\
   & $\C{16}\; (Z/2)$ & (0, 0, 1)\\
   & $\C{17}\; (-Z/2)$ & (0, 0, -1)\\
   \hline\hline
   \end{tabular}
   \end{table}

   We implement 24 single qubit Clifford gates through single-shot protocol in our experiment. To simplify the notation,
   we use the $\mathcal{C}_i$, ($i=0,1,...,23$)  as the names of the Clifford gates with the rotation axes and rotation angles
   shown in the Tab.~\ref{tab1}.
\end{document}